\documentclass[useAMS,usenatbib]{mn2e}
\usepackage{amsmath}
\usepackage{amssymb}
\include{graphicx}

\title[HEIDI: An Automated Process for Light Curves]{HEIDI: An Automated Process for the Identification and Extraction of Photometric Light Curves from Astronomical Images}

\author[M. Todd, H. U. Wallon Pizarro, P. Tanga, D. M. Coward and M. G. Zadnik]{M. Todd$^{1}$\thanks{E-mail:
michael.todd@icrar.org (MT)}, H. U. Wallon Pizarro$^{2}$, P. Tanga$^{3}$, D. M. Coward$^{4}$ and M. G. Zadnik$^{1}$\\
$^{1}$Department of Imaging and Applied Physics, Bldg 301, Curtin University, Kent St, Bentley, WA 6102, Australia\\
$^{2}$IBM Research - Australia, 1060 Hay St, West Perth, WA 6005, Australia\\
$^{3}$Laboratoire Lagrange, UMR7293, Universit\'{e} de Nice Sophia-Antipolis, CNRS, Observatoire de la C\^{o}te d'Azur, BP 4229, \\ \hspace{0.75pt} 06304 Nice Cedex 4, France\\
$^{4}$School of Physics, M013, The University of Western Australia, 35 Stirling Hwy, Crawley, WA 6009, Australia}

\begin{document}

\date{}

\pagerange{\pageref{firstpage}--\pageref{lastpage}} \pubyear{}

\maketitle

\label{firstpage}

\begin{abstract}

The production of photometric light curves from astronomical images is a very time-consuming task, taking several hours or even days. Larger data sets improve the resolution of the light curve, however, the time requirement scales with data volume. The data analysis is often made more difficult by factors such as a lack of suitable calibration sources and the need to correct for variations in observing conditions from one image to another. Often these variations are unpredictable and corrections are based on experience and intuition.

The High Efficiency Image Detection \& Identification (HEIDI) pipeline software rapidly processes sets of astronomical images, taking only a few minutes. HEIDI automatically selects multiple sources for calibrating the images using a selection algorithm that provides a reliable means of correcting for variations between images in a time series. The algorithm takes into account that some sources may intrinsically vary on short time scales and excludes these from being used as calibration sources. HEIDI processes a set of images from an entire night of observation, analyses the variations in brightness of the target objects and produces a light curve all in a matter of minutes.

HEIDI has been tested on three different time series of 50 images each of asteroid 939 Isberga and has produced consistent high quality photometric light curves in a fraction of the usual processing time. The software can also be used for other transient sources, e.g. gamma-ray burst optical afterglows, Gaia transient candidates.

HEIDI is implemented in the programming language Python and processes time series astronomical images in FITS format with minimal user interaction. HEIDI processes up to 1000 images per run in the standard configuration. This limit can be easily increased, with the only real limit being system capacity, e.g. disk space, memory. HEIDI is not telescope-dependent and will process images even in the case that no telescope specifications are provided. HEIDI has been tested on various Linux systems and Linux virtual machines. HEIDI is very portable and extremely versatile with minimal hardware requirements.

\end{abstract}


\section{Introduction}

The process of analysing a time series of astronomical images to measure the variations in brightness of an object over time and to calibrate those variations against reference sources is a very time-consuming task. 
An accurate understanding of the variation in brightness of an object over time, i.e. its photometric light curve, allows modelling of key characteristics of the object (e.g. the shape of an asteroid, the type of a supernova or gamma-ray burst optical afterglow).

The measurement process often varies only little depending on the type of object, whether it is an asteroid tumbling through space, a supernova or the rapidly fading optical afterglow from a gamma-ray burst.
The process usually involves examining the images in the time series for at least one suitable reference source in order to calibrate the brightness. Some processes require the reference sources to have particular characteristics, e.g. a specific spectral class or magnitude.  Other processes may need sources that have been identified as a photometric standard. If more than one calibration source is available, the precision can be improved enormously; however, this is also much more time-expensive.
The process must be repeated for each image and corrections made from one image to another. These corrections can incorporate predictable adjustments such as extinction due to changing air mass. In addition, often the corrections that need to be made in order to obtain a consistent light curve are variable, non-uniform and unpredictable, and are based on experience and intuition. 

The ability to quickly analyse a set of images and to produce a high quality light curve is particularly valuable on automated telescopes, e.g. the fully robotic Zadko Telescope in Gingin, Western Australia \citep{2010PASA...27..331C}. Participation in programmes like the prompt follow-up of gamma-ray burst alerts for the detection of transient optical emissions as part of the T\'{e}lescope \`{a} Action Rapide pour les Objets Transitoires (TAROT) network \citep{2008PASP..120.1298K} and the Gaia Follow-Up Network for Solar System Objects (Gaia-FUN-SSO, http://www.imcce.fr/gaia-fun-sso) \citep{2013PASA...30...14T} results in the creation of data sets containing transient sources that need rapid analysis. 

We have developed a High Efficiency Image Detection \& Identification (HEIDI) pipeline software for processing images and analysing the variations in brightness of optical transients. 
We take the approach that all sources in an image are potential reference sources.
The variations in the brightness of the sources in each image of a time series are assessed automatically. Corrections are made to compensate for these variations while also taking into account that some sources may intrinsically vary on short time scales (e.g. short period variable stars).
To facilitate production of photometric light curves Solar System objects are identified automatically during the image processing.
As a result the automated analysis of an image set and the production of a calibrated light curve is achieved in a matter of minutes. HEIDI thus fulfils an emerging need for the rapid analysis of time series image data.

\section{Description of the automated pipeline software HEIDI}

HEIDI is implemented in the programming language Python (http://www.python.org) for Linux platforms. It processes astronomical images in FITS format and requires only a minimal amount of user-supplied detail. It can also be installed on Linux virtual machines. This makes the software extremely versatile.

Preprocessing tasks such as astrometric calibration and image alignment are accomplished using existing applications as described in \S\ref{section:fitsalign}. 
The main functions of identification and selection of calibration sources, identification of targets and production of light curves are accomplished using newly developed Python code as described in \S\ref{section:correlstar} and \S\ref{section:photometry}. 
HEIDI has been benchmarked on a Pentium class system and a virtual machine, i.e. low-performance, low-cost systems.

\subsection{Astrometric calibration and image alignment} \label{section:fitsalign}

The first step in the image analysis process is to perform the astrometric calibration.
The amount of information contained in FITS image headers depends heavily on the system in place at the observing site. 
While an integrated system would typically produce FITS images which include calibration information in the image headers, some sites may not even include the pointing information. 
We have therefore adopted a standard approach of uploading one image from the set to the astrometry.net server in order to obtain a new astrometric calibration \citep{2010AJ....139.1782L}.
By using the blind astrometric calibration system provided by astrometry.net, HEIDI thus removes any reliance on the images having already had astrometric calibrations or having to verify the pointing information, orientation or scale.

After obtaining the astrometric calibration, HEIDI assumes that all the images in the time series have the same pointing and updates the FITS image headers. The World Coordinate System (WCS) information from astrometry.net is included in the headers of the updated FITS images. The images are then processed through SExtractor \citep{1996A&AS..117..393B} in order to build a catalogue of all objects found in the image set. The SExtractor catalogue is subsequently processed through SCAMP \citep{2006ASPC..351..112B} which produces image headers that are ready to be used by SWarp \citep{2002ASPC..281..228B}. SWarp aligns the images and includes the appropriate header information from any translation or rotation applied by SWarp during the alignment process. Most of the processing time is taken up with the SExtractor-SCAMP-SWarp workflow. On a single-CPU Pentium-class system, the astrometric calibration and image alignment stage for a set of 50 images takes about 10~min (vs. 1.25~min for the automated production of the photometric light curve by HEIDI, \S\ref{section:correlstar}, \S\ref{section:photometry}).

\subsection{Correlation of sources} \label{section:correlstar}

In the correlation stage the images from the astrometric calibration and alignment stage are reprocessed through SExtractor to produce catalogues of all the sources that have been detected across the entire image set.
These catalogues are correlated by HEIDI to identify the sources which are common to every image in the time series. These common sources are the candidate reference sources for the photometric calibration of the image set.

\begin{figure}
\includegraphics{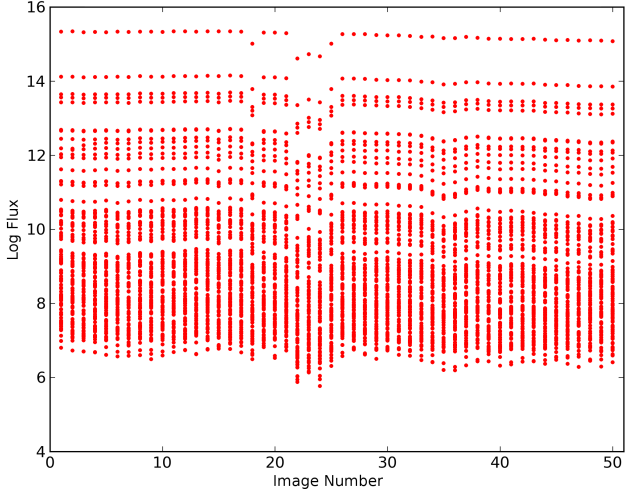}
\caption{\label{fig:figure1}The selected stars from observations of asteroid 939 Isberga show a variation in the log flux values across the time series. As this variation appears to affect every source in a consistent manner it can be analysed and a correction applied. }
\end{figure}

HEIDI calculates the natural logarithm of the flux value for each candidate source as a representation of relative (instrumental) magnitude. We refer to the natural logarithm of the flux hereafter simply as log flux. Examination of the log flux value for each candidate source may show some variation across the image set. HEIDI tests the log flux variation of each candidate source and rejects sources as candidates if the variation is greater than a predetermined threshold. An examination of the log flux values for the candidate sources in a time series of 50 images for asteroid 939 Isberga\footnote{Three time series of 50 images each were taken by the first author (M. Todd) in November 2011 at the Zadko Telescope in Gingin, Western Australia.} shows a variation common to all candidate sources (Figure~\ref{fig:figure1}). As this variation appears to affect every source in a consistent manner, we conclude that this variation can be analysed and corrected.  

For a given time series of $N$ images, HEIDI finds all sources that are common to all $N$ images in that time series. These $M$ sources are selected as candidates for photometric calibration and are tested for variability over the time series. 

STEP 1: For the $N$ images and $M$ sources, HEIDI reads the flux values $\phi_{ki}$, $k\in\{1,\ldots,N\}$, $i\in\{1,\ldots,M\}$ from the SExtractor catalogues and calculates the natural logarithm of the flux. We will denote the natural logarithm of the flux as $\varphi_{ki}$, $k\in\{1,\ldots,N\}$, $i\in\{1,\ldots,M\}$ and refer to it hereafter as log flux. The initial log flux set is hence $\big(\varphi_{ki}\big)_{\substack{k\in\{1,\ldots,N\}\\ i\in\{1,\ldots,M\}}}$, $\varphi_{ki} = \ln\phi_{ki}$ and contains $N\cdot{M}$ values.

STEP 2:
For each image $k\in\{1,\ldots,N\}$ in the time series, HEIDI calculates $M$ log flux means: For each source $i\in\{1,\ldots,M\}$, the source $i$ is excluded (denoted by $\overline{i}$) and the log flux mean of the remaining sources $j\in\{1,\ldots,M\}\setminus\{i\}$ is determined:

\begin{equation} \label{eqn:mean_log_flux}
\begin{split}
  \left\langle{\varphi_{k\overline{i}}}\right\rangle &= \left\langle{\varphi_{kj}}\right\rangle_{j\in\{1,\ldots,M\}\setminus\{i\}} \\ 
                                                     &= \frac{1}{M-1} \sum_{\substack{j=1\\j \neq i}}^M \varphi_{kj}, \quad k\in\{1,\ldots,N\}.
\end{split}
\end{equation}

STEP 3:
HEIDI determines the global maximum of the log flux: 

\begin{equation} \label{eqn:glob_max_log_flux}
  \varphi_{max} = \max\limits_{k,i}\varphi_{ki} = \max\limits_{\substack{k\in\{1,\ldots,N\}\\ i\in\{1,\ldots,M\}}}\varphi_{ki}. 
\end{equation}

For all $k\in\{1,\ldots,N\}$, $i\in\{1,\ldots,M\}$ the difference between the global maximum log flux $\varphi_{max}$ and the log flux means $\left\langle{\varphi_{k\overline{i}}}\right\rangle$ (determined in STEP 2) are calculated:

\begin{equation} \label{eqn:corr_factor}
\begin{split}
  & \Delta\varphi_{ki} = \varphi_{max} - \left\langle{\varphi_{k\overline{i}}}\right\rangle, \\
  & \qquad \qquad \qquad k\in\{1,\ldots,N\}, i\in\{1,\ldots,M\}.
\end{split}
\end{equation}

This will be our correction factor.

STEP 4:
To smooth out the variations across the image set, the log flux values $\varphi_{ki}$, $k\in\{1,\ldots,N\}$, $i\in\{1,\ldots,M\}$ are adjusted, yielding an adjusted log flux set: 

\begin{equation} \label{eqn:adj_log_flux}
\begin{split}
   & \big(\varphi^\ast_{ki}\big)_{\substack{k\in\{1,\ldots,N\}\\ i\in\{1,\ldots,M\}}}, \varphi^\ast_{ki} = \varphi_{ki} + \Delta\varphi_{ki}, \\ 
   & \qquad \qquad \qquad k\in\{1,\ldots,N\}, i\in\{1,\ldots,M\}.
\end{split}
\end{equation}

STEP 5:
For each source $i\in\{1,\ldots,M\}$ HEIDI determines the adjusted log flux mean, the maximum adjusted log flux and the minimum adjusted log flux of this source across all $N$ images:

\begin{equation} \label{eqn:mean_adj_log_flux}
  \left\langle{\varphi^\ast_{ki}}\right\rangle_{k\in\{1,\ldots,N\}} = \frac{1}{N} \sum_{k=1}^N \varphi^\ast_{ki}, \quad i\in\{1,\ldots,M\},
\end{equation}

\begin{equation} \label{eqn:max_adj_log_flux}
  \max\limits_{k}\varphi^\ast_{ki} = \max\limits_{k\in\{1,\ldots,N\}}\varphi^\ast_{ki}, \quad i\in\{1,\ldots,M\},
\end{equation}

\begin{equation} \label{eqn:min_adj_log_flux}
  \min\limits_{k}\varphi^\ast_{ki} = \min\limits_{k\in\{1,\ldots,N\}}\varphi^\ast_{ki}, \quad i\in\{1,\ldots,M\}.
\end{equation}

STEP 6:
For each source $i\in\{1,\ldots,M\}$ HEIDI tests whether the deviation of the adjusted log flux values $\big(\varphi^\ast_{ki}\big)_{\substack{k\in\{1,\ldots,N\}\\ i\in\{1,\ldots,M\}}}$ from the adjusted log flux mean $\left\langle{\varphi^\ast_{ki}}\right\rangle_{k\in\{1,\ldots,N\}}$ is within a predetermined tolerance $tol$. This test rejects sources that are varying in brightness on short time scales (e.g. short period variable stars) or are otherwise not varying in accordance with the other candidates in the set (e.g. edge effects or vignetting). Only sources that pass this test will remain in the candidate list:

\begin{equation}\label{eqn:tol}
\begin{split}
  & |\max\limits_{k}\varphi^\ast_{ki} - \left\langle{\varphi^\ast_{ki}}\right\rangle_{k\in\{1,\ldots,N\}}| \le tol \quad \wedge \\
  & |\min\limits_{k}\varphi^\ast_{ki} - \left\langle{\varphi^\ast_{ki}}\right\rangle_{k\in\{1,\ldots,N\}}| \le tol, \quad i\in\{1,\ldots,M\}.
\end{split}
\end{equation}

Sources that do not pass this test will be excluded from the candidate list. If $X$ is the number of excluded sources then the reduced candidate list consists of $M - X$ candidates. This concludes the first pass and reduces the original candidate list 
$\big(\varphi_{ki}\big)_{\substack{k\in\{1,\ldots,N\}\\ i\in\{1,\ldots,M\}}}$ 
to
$\big(\varphi'_{ki}\big)_{\begin{subarray}{l}k\in\{1,\ldots,N\}\\i\in\{1,\ldots,M\}\setminus\{i_1,\ldots,i_X\}\end{subarray}}$:

\begin{equation}\label{eq:source_reduction}
\begin{split}
  & \big(\varphi'_{ki}\big)_{\begin{subarray}{l}k\in\{1,\ldots,N\}\\ i\in\{1,\ldots,M\}\setminus\{i_1,\ldots,i_X\}\end{subarray}} = \\
  & \qquad \qquad \big(\varphi_{ki}\big)_{\substack{k\in\{1,\ldots,N\}\\ i\in\{1,\ldots,M\}}} \setminus \big(\varphi_{ki}\big)_{\begin{subarray}{l}k\in\{1,\ldots,N\}\\ i\in\{i_1,\ldots,i_X\}\end{subarray}}, \\
  & \qquad \qquad \qquad \qquad \{i_1,\ldots,i_X\}\subseteq\{1,\ldots,M\}.
\end{split}
\end{equation}

STEP 7:
Repeat STEP 2 to STEP 6 for the reduced candidate list $\big(\varphi'_{ki}\big)_{\begin{subarray}{l}k\in\{1,\ldots,N\}\\ i\in\{1,\ldots,M\}\setminus\{i_1,\ldots,i_X\}\end{subarray}}$.

This process of filtering variable sources is applied iteratively. The tolerance $tol$ for the deviation test of the log flux (STEP 6) is 0.5, 0.2 and 0.1 for the first, second and third pass, respectively. We found that additional iterations with finer tolerances do not significantly improve the result as one reaches the limit of measurement precision. The candidate stars that survive this filtering process are used in the photometric calibration of the image set for producing the light curve for the target as described in \S\ref{section:photometry}. In addition, the candidates that fail the test are recorded for further study.

\begin{figure}
\includegraphics{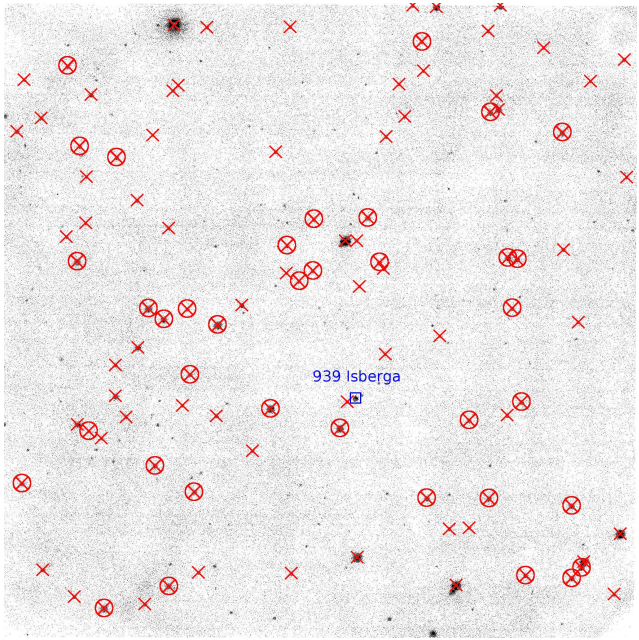}
\caption{\label{fig:figure2}Sample image from the time series for asteroid 939 Isberga. This image is centred on right ascension $01^h~08^m~32.58^s$, declination $+11\degr~45'~48.43''$ (J2000) and has a field of view of $\approx20\text{x}20$~arcmin. Sources that are common to all images in the time series are indicated by a cross $(\times)$. These sources are tested for their suitability to be used as calibration sources. Circles $(\otimes)$ indicate those sources that have been selected as calibration sources. The position of the target, asteroid 939 Isberga is indicated by a square ($\square$). }
\end{figure}

Figure~\ref{fig:figure2} shows a sample image from a time series of $N=50$ images for asteroid 939 Isberga. This image contains exactly $M=100$ sources that are common to all images in the time series, marked with a cross $(\times)$. These sources are selected as candidate sources and are tested for their suitability for use as calibration sources. Circles $(\otimes)$ indicate the reduced set of 37 sources that were found to be suitable for use as calibration sources by HEIDI. The position of the target, asteroid 939 Isberga is indicated by a square ($\square$).

On a single-CPU Pentium-class system the correlation stage for a time series of 50 images takes only about 1~min.

\subsection{Identification of target sources and production of the light curves} \label{section:photometry}

HEIDI has been designed to facilitate the production of photometric light curves for Solar System objects (target sources). 
Following the selection of calibration sources as described in \S\ref{section:correlstar}, the target sources are identified and selected. 
HEIDI uses the Institut de M\'{e}canique C\'{e}leste et de Calcul des \'{E}ph\'{e}m\'{e}rides (IMCCE) Sky Body Tracker (SkyBoT) Cone-Search and Resolver services \citep{2006ASPC..351..367B} to quickly identify the known Solar System objects in the time series, our target sources.
HEIDI processes the time series very quickly: For a set of 50 images it takes HEIDI only approximately 15~s to identify one target source, to test for its presence in the image set, to compile the list of the corresponding flux values and to produce the photometric light curve for this object, i.e. $\approx0.3$~s per object per image.
As a consequence we have implemented HEIDI to produce light curves for all of the objects found in the time series rather than just the primary target.

The IMCCE SkyBoT Cone-Search service identifies which Solar System objects lie within a specified field of view at a given epoch: HEIDI calls the Cone-Search service and obtains a list of the objects that may be found in the time series by providing the centre coordinates in right ascension and declination and the epoch of observation. 

For each object returned by the Cone-Search service, HEIDI calls the IMCCE SkyBoT Resolver service to obtain accurate coordinates: HEIDI provides the name of the object and the epoch for the first image in the time series to the Resolver service. If known, HEIDI also provides the International Astronomical Union (IAU) Observatory Code for the site at which the images were recorded. The Resolver service then returns accurate coordinates for the specified object. If the IAU Observatory Code is provided then the Resolver service returns coordinates in the topocentric frame, otherwise the coordinates are given for the geocentric frame. 

To determine the object coordinates in the remaining images of the time series, HEIDI finds the record for each object in the IAU Minor Planet Center Orbit (MPCORB) database (http://www.minorplanetcenter.net). 
The coordinates for each object are computed from the orbital elements in the MPCORB database on a per-image basis across the time series. These computations are much faster than obtaining coordinates from the Resolver service. It takes only a fraction of a second to calculate the coordinates for an object for all images in the time series, compared with several seconds per image when retrieved from the Resolver service. 
However, due to rounding errors in the computation, the coordinates provided by the Resolver service may be more accurate than the ones computed by HEIDI. 
HEIDI will thus determine the difference between the coordinates provided by the Resolver service and the ones computed locally and apply this offset as a correction to all computed coordinates across all images.

The respective coordinates in each image of the time series are then tested to determine the presence of the target sources. To minimise false detections without introducing false non-detections, this is done in a two-step process: In a first step, an area of 10~x~10~pixels around the coordinates of the target is tested for the presence of a source; if no source is found, the area is expanded to 15~x~15~pixels and the test repeated.
It is not possible to determine whether a failure to detect a target source is due to the faintness of the object or an error in the position calculation. We thus assume that the primary target for the measurement and production of a light curve has a well-known orbit and also a good signal-to-noise ratio in the time series.  

\begin{figure}
\includegraphics{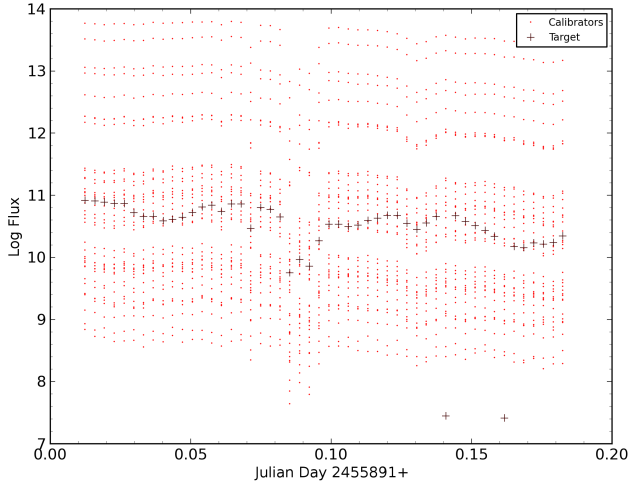}
\caption{\label{fig:figure3}Uncorrected log flux measurements from calibration sources $(\cdot)$ and target $(+)$ in the time series for asteroid 939 Isberga. The calibration sources can clearly be seen to vary in accordance with each other. The target, asteroid 939 Isberga, follows a similar trend. The light curve is not readily apparent. }
\end{figure}

If the object is detected in the images of the time series, the log flux of the object is recorded for each image. Plotting the log flux of the object and the log flux of the calibration sources without correcting for the variations between images in the time series does not give a clear indication of the light curve of an object. Figure~\ref{fig:figure3} shows the uncorrected log flux measurements from calibration sources and the target in a time series for asteroid 939 Isberga. 

The log flux mean of the calibration sources identified in \S\ref{section:correlstar} is calculated for each image (similar to Equation~\ref{eqn:mean_log_flux} but this time over the entire candidate source set):
\begin{equation}\label{eqn:mean_log_flux2}
  \left\langle{\varphi_{ki}}\right\rangle_{i\in\{1,\ldots,M\}} = \frac{1}{M} \sum_{i=1}^M \varphi_{ki}, \quad k\in\{1,\ldots,N\}.
\end{equation}

The global log flux mean is also determined: 
\begin{equation}\label{eqn:global_log_flux_mean}
\begin{split}
  \left\langle{\varphi_{ki}}\right\rangle_{\substack{k\in\{1,\ldots,N\}\\i\in\{1,\ldots,M\}}} & = \frac{1}{NM} \sum_{k=1}^N \sum_{i=1}^M \varphi_{ki} \\
                                                                                              & \stackrel{Eq. \ref{eqn:mean_log_flux2}}{=} \frac{1}{N} \sum_{k=1}^N \left\langle{\varphi_{ki}}\right\rangle_{i\in\{1,\ldots,M\}}.
\end{split}
\end{equation}

This is used for reference and correction of variations across the time series.

For each image $k\in\{1,\ldots,N\}$ in the time series, the log flux values of the calibration sources are adjusted by adding the difference between the global log flux mean and the log flux mean for that image. For all $k\in\{1,\ldots,N\}$, $i\in\{1,\ldots,M\}$ the difference between the global log flux mean $\left\langle{\varphi_{ki}}\right\rangle_{\substack{k\in\{1,\ldots,N\}\\i\in\{1,\ldots,M\}}}$ and the log flux mean $\left\langle{\varphi_{ki}}\right\rangle_{i\in\{1,\ldots,M\}}$ is calculated:

\begin{equation}\label{eqn:corr_factor2}
\begin{split}
  & \Delta\varphi_{k} = \left\langle{\varphi_{ki}}\right\rangle_{\substack{k\in\{1,\ldots,N\}\\i\in\{1,\ldots,M\}}} - \left\langle{\varphi_{ki}}\right\rangle_{i\in\{1,\ldots,M\}}, \\
  & \qquad \qquad \qquad \qquad \qquad \qquad k\in\{1,\ldots,N\}.
\end{split}
\end{equation}

For each image $k\in\{1,\ldots,N\}$ in the time series, the log flux value of the target object $\varphi_{kt}$, $k\in\{1,\ldots,N\}$ (where $t$ denotes the target source) is adjusted by the same correction factor $\Delta\varphi_{k}$. To smooth out the background variations across the image set, the log flux values $\varphi_{kt}$, $k\in\{1,\ldots,N\}$ for the target object $t$ are adjusted:

\begin{equation} \label{eqn:adj_log_flux_target}
   \varphi^\ast_{kt} = \varphi_{kt} + \Delta\varphi_{k}, \quad k\in\{1,\ldots,N\}.
\end{equation}

\begin{figure}
\includegraphics{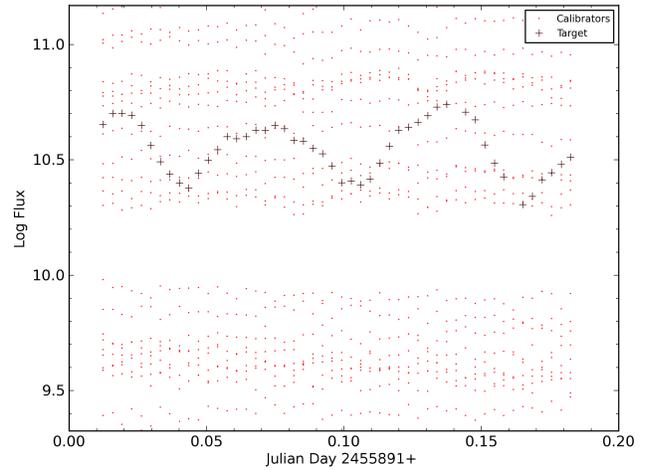}
\caption{\label{fig:figure4}Corrected log flux measurements for calibration sources $(\cdot)$ and target $(+)$ in the time series for asteroid 939 Isberga. The variation in the calibration sources is minimal $(<\pm0.05)$. The light curve of the target is clearly apparent. }
\end{figure}

Applying this correction to the time series for asteroid 939 Isberga shows a light curve that is readily apparent (Figure~\ref{fig:figure4}). This light curve for asteroid 939 Isberga produced by HEIDI clearly shows a primary period consistent with the expected rotation period of 2.9173~h (0.12155~d) \citep{2008MPBu...35....9M}.

\begin{figure}
\includegraphics{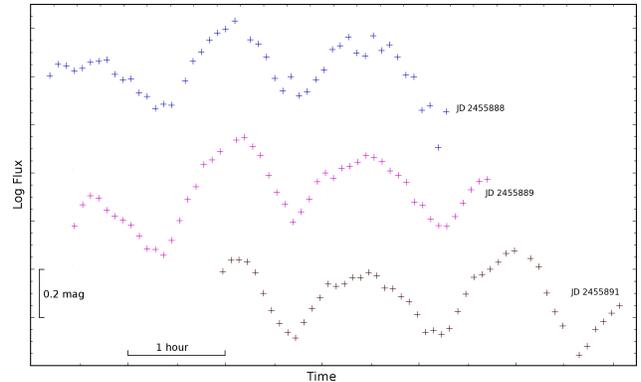}
\caption{\label{fig:figure5}The light curves produced by HEIDI for asteroid 939 Isberga using time series that were obtained on three separate dates show the consistency of the results. It can be easily seen that the three curves follow a very similar pattern. A primary periodicity consistent with the expected rotation period of 2.9173~h is evident. }
\end{figure}

The consistency of the results achieved by HEIDI is further demonstrated by extending the analysis to multiple time series for asteroid 939 Isberga that were obtained on three separate dates. The light curves determined by HEIDI for these three time series are shown superimposed in Figure~\ref{fig:figure5}. It can be easily seen that the three curves follow a very similar pattern.

\begin{figure}
\includegraphics{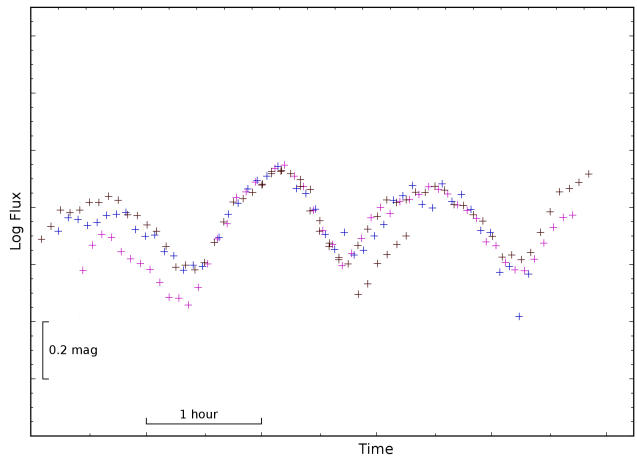}
\caption{\label{fig:figure6}Light curves produced by HEIDI for asteroid 939 Isberga from time series obtained on three separate dates (22, 23 and 25 November 2011) overlaid on one another to show that both the period and the amplitude of each curve is consistent with the period and amplitude of the other curves. }
\end{figure}

These curves can also be overlaid on one another to show that both the period and the amplitude are consistent (Figure~\ref{fig:figure6}). 

\section{Future developments}
The current version (as of November 2013) has good astrometric accuracy but performs only relative photometry, i.e. does not produce calibrated magnitudes. We plan to add the capability for the production of light curves calibrated to a standard photometric system.

The preprocessing tasks of astrometric calibration and image alignment described in \S\ref{section:fitsalign} are independent of the type of object being studied. 
The candidate sources that are rejected by the filtering process described in \S\ref{section:correlstar} warrant further study to determine whether they were excluded due to some intrinsic variability. 
These excluded sources can include non-moving sources that have characteristic light curves, e.g. variable stars, supernovae and gamma-ray burst optical afterglows. 
In the next version of HEIDI, we will include the ability to also produce photometric light curves for all the excluded objects as a standard function.

Uncertainties on the produced light curves can be inferred from the dispersion of the magnitudes of the calibration sources and will be used to include error bars in the output.
If the target source is near, i.e. within 2 to 3 arcsec, a star of similar or greater brightness, it is possible that the star is selected instead of the target source. The next version of HEIDI will include a step to automatically exclude detections in which the position information falls outside a predetermined tolerance. In addition, an algorithm to detect moving objects will be implemented for targets with less accurate position information or unknown targets.

\section{Conclusions}

The High Efficiency Image Detection \& Identification (HEIDI) pipeline software rapidly processes sets of astronomical images. It is very fast and reliable. HEIDI takes a set of images from a night of observing and produces a light curve in a matter of minutes.

HEIDI is implemented in the programming language Python. It has been tested on various Linux systems and Linux virtual machines. It has minimal hardware requirements and the installation process is relatively straightforward. 
HEIDI runs in virtual machines without impacting performance and can thus be installed on non-Linux systems (if a Linux virtual machine is available). 
The High Efficiency Image Detection \& Identification (HEIDI) pipeline software we developed has a small footprint yet is very powerful and extremely versatile.

Given programmes like the prompt follow-up of gamma-ray burst alerts for the detection of transient optical emissions and the Gaia Follow-Up Network for Solar System Objects, which require prompt follow-up and rapid analysis of time series data, HEIDI will be very useful and provide a service and a speed that have not been available.

Testing HEIDI on multiple fairly large time series for asteroid 939 Isberga yielded extremely satisfying results. The light curves produced using time series obtained on three separate dates showed consistency with the three curves following a very similar pattern.

HEIDI will be installed at the Zadko Telescope in Gingin, Western Australia. This will allow us to analyse very rapidly all the images in the data archive of the Zadko Telescope and search for interesting variable sources. It will also allow prompt analysis and rapid issuing of results for time critical events such as gamma-ray burst optical afterglow emissions.

The HEIDI package is available from the first author on request.

\section*{Acknowledgments}

MT thanks Fr\'{e}d\'{e}ric Vachier for assistance with photometry.
MT thanks Andrew Williams for assistance with processing FITS image files. 
The work reported on in this publication has been supported by the European Science Foundation (ESF), in the framework of the GREAT Research Networking Programme.
DMC is supported by an Australian Research Council Future Fellowship.

\label{lastpage}

\end{document}